\documentclass[journal]{IEEEtran}
\usepackage{graphicx} 
\usepackage{booktabs}
\usepackage[table,xcdraw]{xcolor}
\usepackage{url}
\usepackage{tabularx}
\usepackage{xcolor}
\usepackage{booktabs}
\usepackage{adjustbox}
\usepackage{float}
\usepackage{amsmath}
\usepackage{graphicx}
\usepackage{booktabs}
\usepackage{multirow}
\usepackage{cite}
\usepackage{subcaption}
\usepackage{graphicx}
\usepackage{algorithm}
\usepackage{algpseudocode}
\usepackage{multirow}

\ifCLASSINFOpdf

\else

\fi

\begin{document}

\title{Vision Transformer for Intracranial Hemorrhage  Classification in CT Scans Using an Entropy-Aware Fuzzy Integral Strategy for Adaptive Scan-Level Decision Fusion}

\author{Mehdi Hosseini Chagahi, Md. Jalil Piran, Niloufar Delfan, Behzad Moshiri, Jaber Hatam Parikhan

\thanks{(Corresponding Authors: B. Moshiri and M. J. Piran)}
\thanks{M. H. Chagahi and N. Delfan are with the School of Electrical and Computer Engineering, College of Engineering, University of Tehran, Tehran, Iran, (e-mail: mhdi.hoseini@ut.ac.ir;
niloufar.delfan@ut.ac.ir )}

\thanks{B. Moshiri is with the School of Electrical and Computer Engineering, College of Engineering, University of Tehran, Tehran, Iran and the Department of Electrical and Computer Engineering University of Waterloo,
Waterloo, Canada, (e-mail: moshiri@ut.ac.ir)}

\thanks{M. J. Piran is with the Department of Computer Science and Engineering, Sejong University, Seoul 05006, South Korea, (e-mail: piran@sejong.ac.kr)}}

\maketitle

\begin{abstract}
Intracranial hemorrhage (ICH) is a critical medical emergency caused by the rupture of cerebral blood vessels, leading to internal bleeding within the skull. Accurate and timely classification of hemorrhage subtypes is essential for effective clinical decision-making. To address this challenge, we propose an advanced pyramid vision transformer (PVT)-based model, leveraging its hierarchical attention mechanisms to capture both local and global spatial dependencies in brain CT scans. Instead of processing all extracted features indiscriminately, A SHAP-based feature selection method is employed to identify the most discriminative components, which are then used as a latent feature space to train a boosting neural network, reducing computational complexity. We introduce an entropy-aware aggregation strategy along with a fuzzy integral operator to fuse information across multiple CT slices, ensuring a more comprehensive and reliable scan-level diagnosis by accounting for inter-slice dependencies. Experimental results show that our PVT-based framework significantly outperforms state-of-the-art deep learning architectures in terms of classification accuracy, precision, and robustness. By combining SHAP-driven feature selection, transformer-based modeling, and an entropy-aware fuzzy integral operator for decision fusion, our method offers a scalable and computationally efficient AI-driven solution for automated ICH subtype classification.
\end{abstract}


\IEEEpeerreviewmaketitle

\section{Introduction}
ICH is a type of stroke characterized by bleeding within the brain tissue or surrounding spaces \cite{puy2023intracerebral}. This condition is life-threatening, requiring immediate medical attention to prevent severe neurological damage or death. Despite accounting for only 10–15\% of all stroke cases globally, ICH is associated with high mortality and morbidity rates \cite{renedo2024burden}. Approximately 66\% of all deaths from neurological diseases are linked to brain hemorrhages, with nearly half of these fatalities occurring within the first 24 hours of the hemorrhagic event \cite{smith2024cavernous}. Timely detection and classification of ICH subtypes, such as intraparenchymal hemorrhage (IPH), intraventricular hemorrhage (IVH), epidural hemorrhage (EDH), subarachnoid hemorrhage (SAH), and subdural hemorrhage (SDH), are crucial for effective treatment and improved patient outcomes \cite{neethi2024comprehensive}.

Accurate diagnosis of ICH remains challenging because of several factors. First, interpreting computed tomography (CT) scans requires substantial expertise, as different hemorrhage types exhibit diverse imaging patterns \cite{ahmed2023systematic, bayoudh2024survey}. Even experienced radiologists may struggle with borderline cases or complex presentations, leading to interobserver variability \cite{li2024code}. Second, the rapid decision-making required in emergency settings increases the likelihood of diagnostic errors, especially when dealing with a high volume of patients \cite{wang2025cross,chagahi2024cardiovascular, chagahi2024enhancing}. Third, existing conventional methods, including rule-based systems and traditional computer-aided diagnosis (CAD) tools, often lack robustness in differentiating similar-appearing hemorrhages \cite{nizarudeen2024comparative}. These challenges highlight the need for automated, AI-driven solutions that can assist clinicians in achieving higher diagnostic accuracy, consistency, and efficiency \cite{mansour2023artificial, seners2023role}.

Recent advancements in machine learning (ML) have significantly improved medical image analysis, offering data-driven, automated approaches for disease detection and classification \cite{delfan2024ai, iqbal2024hybrid, ding2024fmdnn}. Deep learning (DL) models, particularly convolutional neural networks (CNNs) and transformer-based architectures, have showed remarkable success in feature extraction, segmentation, and classification of complex medical images, including brain CT scans \cite{ren2025conv, zhang2024deep, chagahi2024ai}.

\begin{figure*}[t!]
    \centering
    \begin{subfigure}{0.19 \textwidth}
        \includegraphics[width=\linewidth]{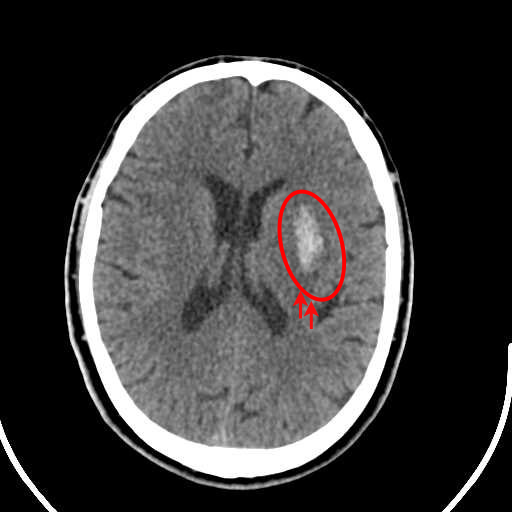}
        \subcaption{} 
    \end{subfigure}
    \begin{subfigure}{0.19 \textwidth}
        \includegraphics[width=\linewidth]{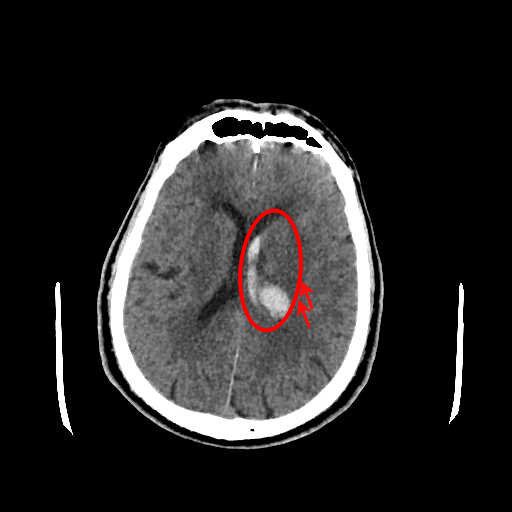}
        \subcaption{} 
    \end{subfigure}
    \begin{subfigure}{0.19 \textwidth}
       \includegraphics[width=\linewidth]{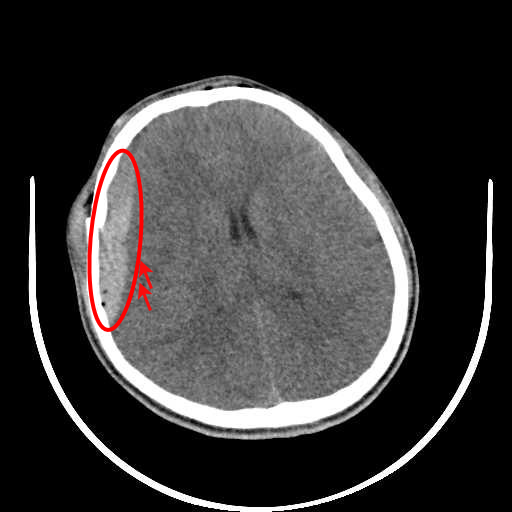}
       \subcaption{} 
    \end{subfigure}
        \begin{subfigure}{0.19 \textwidth}
       \includegraphics[width=\linewidth]{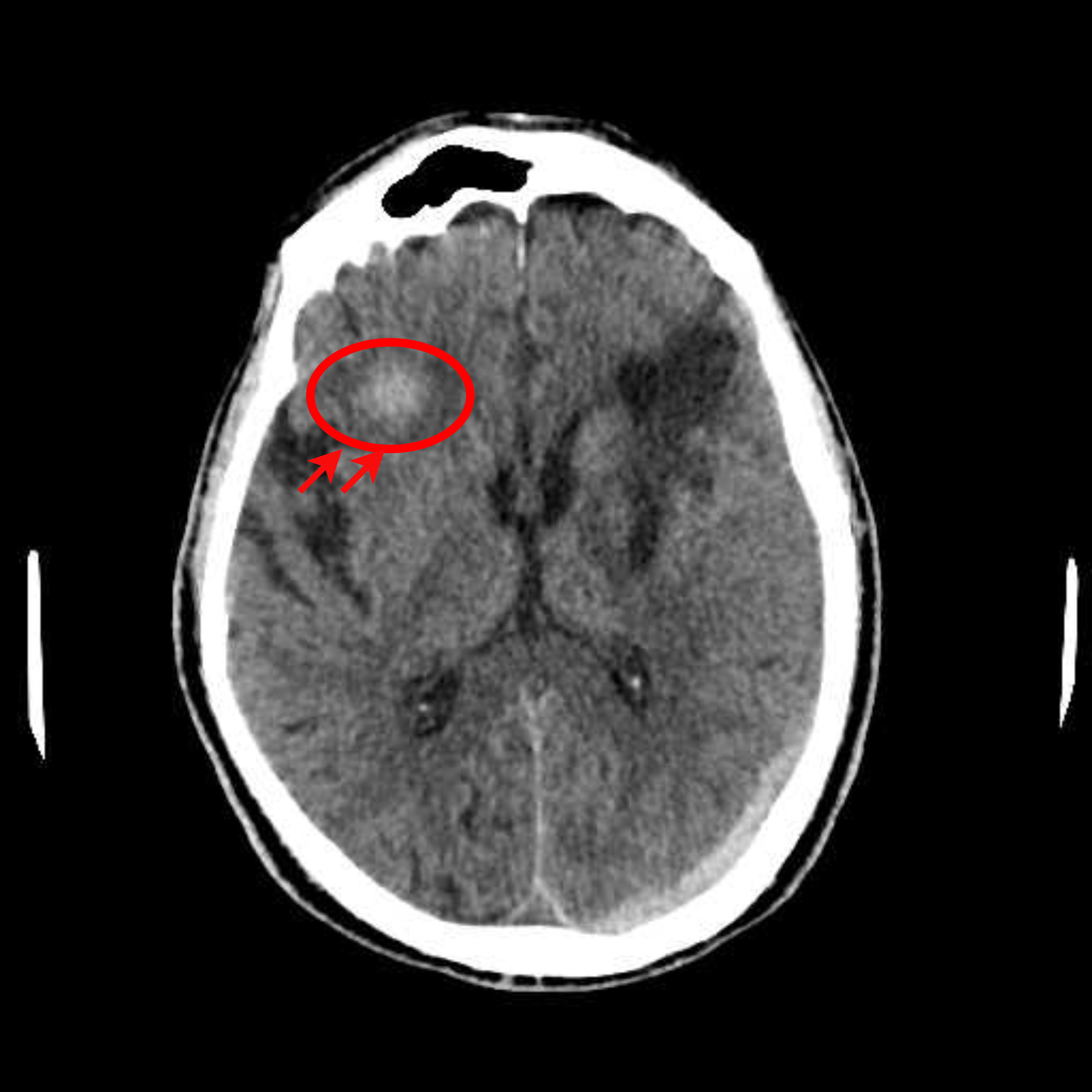}
       \subcaption{} 
    \end{subfigure}
        \begin{subfigure}{0.19 \textwidth}
       \includegraphics[width=\linewidth]{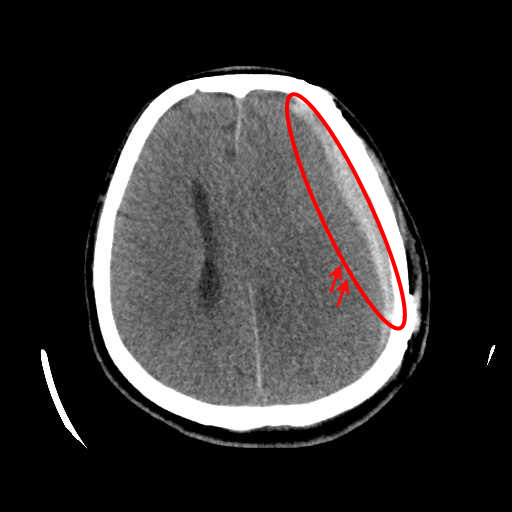}
       \subcaption{} 
    \end{subfigure}
    \caption{Illustrative examples of different types of brain hemorrhages in CT scans. The hemorrhagic regions are highlighted in red. (a) IPH, (b) IVH, (c) EDH, (d) SAH, (e) SDH.}
    \label{labeling}

\end{figure*}

Despite the remarkable advancements in deep learning for ICH detection and classification, existing methods still face critical limitations. CNNs and attention-based models struggle with capturing global spatial dependencies in CT scan images, limiting their ability to generalize across different ICH subtypes. While some approaches incorporate multiple CT slices, they often lack an effective scan-level aggregation strategy, leading to inconsistent predictions across different slices. Current feature selection methods fail to systematically prioritize the most discriminative features, resulting in models that process redundant or non-informative data, increasing computational complexity without improving classification performance. To address these gaps, we propose a vision transformer (PVT)-based model with shapley additive explanations (SHAP)-driven feature selection and an entropy-aware fuzzy integral operator for scan-level decision fusion, significantly enhancing both reliability and diagnostic accuracy.
This study introduces the following key contributions.
\begin{itemize}
    \item Comprehensive Dataset Collection and Expert Labeling: A novel dataset of brain CT scans was collected from two medical centers, covering diverse cases of ICH. All scans were manually labeled by board-certified neurosurgeons using a hierarchical annotation process to ensure high-quality ground truth labels. This expert-driven approach enhances classification reliability and model performance.
    \item Optimized Preprocessing of CT Scan Images: Input images undergo preprocessing techniques, including removal of non-homogeneous color regions and resizing images to focus on the primary brain tissue. This process uses background removal and binary masking techniques to ensure that only the relevant brain regions are kept, improving feature extraction accuracy and reducing computational complexity.
    \item Vision Transformer for ICH Type Classification: A PVT-based model is employed to capture both local and global spatial dependencies in brain CT scans, leading to improved multi-class classification performance compared to CNN-based models.
    \item SHAP-Based Feature Selection for Boosting Neural Network: Instead of processing all extracted features indiscriminately, SHAP analysis applies to identify the most discriminative components, which are then used as a latent feature space for training a boosting neural network, ensuring computational efficiency and model interpretability.
    \item Entropy-Aware Fuzzy Integral Operator for Scan-Level Decision Fusion: A fuzzy integral strategy with an entropy-aware weighting mechanism is introduced to aggregate multiple CT slices, resulting in a more reliable and robust scan-level classification.
\end{itemize}
The rest of this paper is structured: Section \ref{Dataset} describes the dataset collection process, annotation procedures, and preprocessing techniques applied to enhance CT scan quality. Section \ref{Methodology} details the proposed framework, including feature extraction, SHAP-based feature selection, and the entropy-aware fuzzy integral strategy for scan-level fusion. The experimental results and performance evaluation of different models are presented in Section \ref{Findings}. Section \ref{Discussion} discusses our approach to existing methodologies and includes potential directions for future research as a subsection. Finally, Section \ref{Conclusion} summarizes the key contributions and findings of this study.

\begin{figure*}[t!]
    \centering
    \begin{subfigure}{0.19 \textwidth}
        \includegraphics[width=\linewidth]{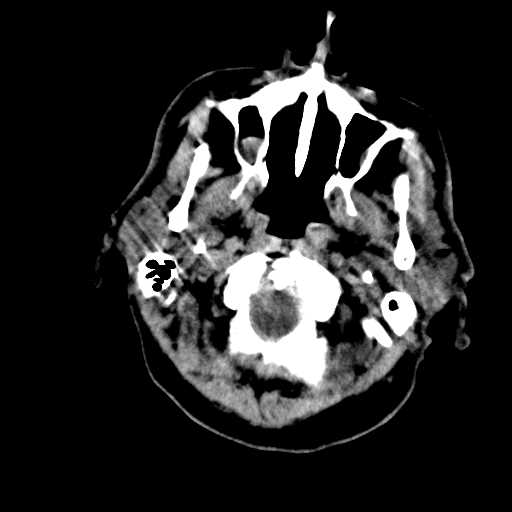}
    \end{subfigure}
    \begin{subfigure}{0.19 \textwidth}
        \includegraphics[width=\linewidth]{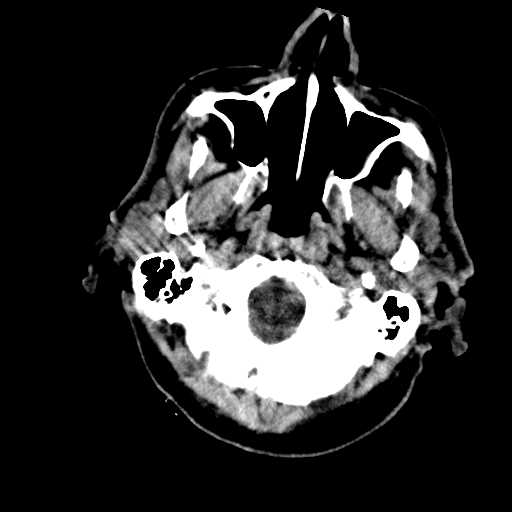}
    \end{subfigure}
    \begin{subfigure}{0.19 \textwidth}
        \includegraphics[width=\linewidth]{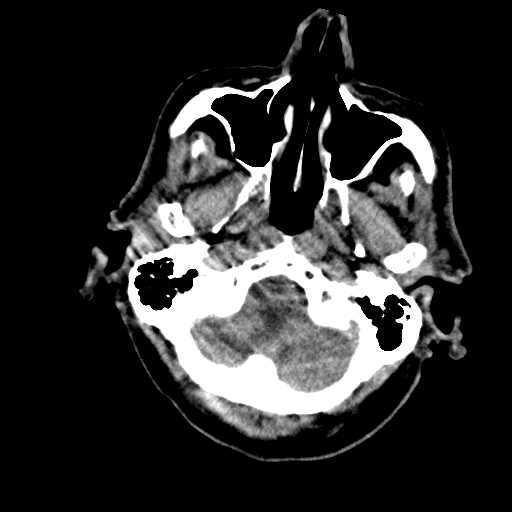}
    \end{subfigure}
    \begin{subfigure}{0.19 \textwidth}
        \includegraphics[width=\linewidth]{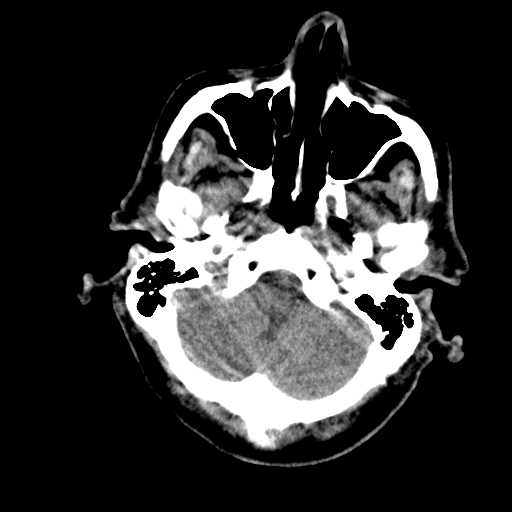}
    \end{subfigure}
    \begin{subfigure}{0.19 \textwidth}
        \includegraphics[width=\linewidth]{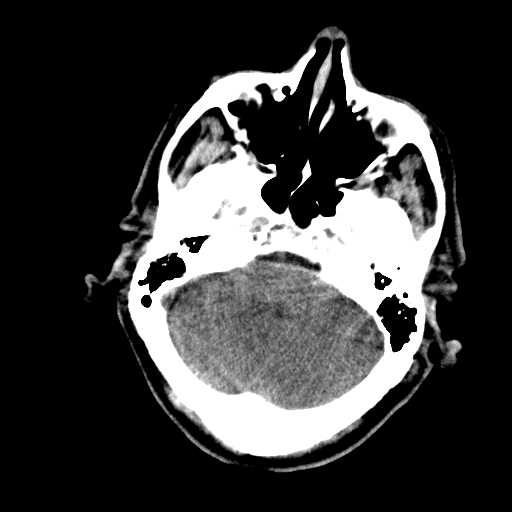}
    \end{subfigure}

    \vspace{0.1cm}

    \begin{subfigure}{0.19 \textwidth}
        \includegraphics[width=\linewidth]{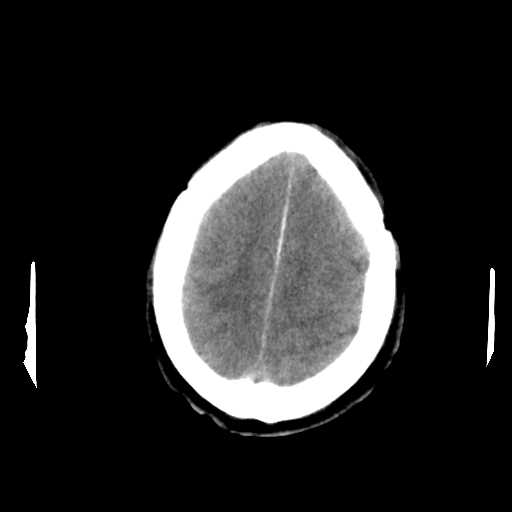}
    \end{subfigure}
    \begin{subfigure}{0.19 \textwidth}
        \includegraphics[width=\linewidth]{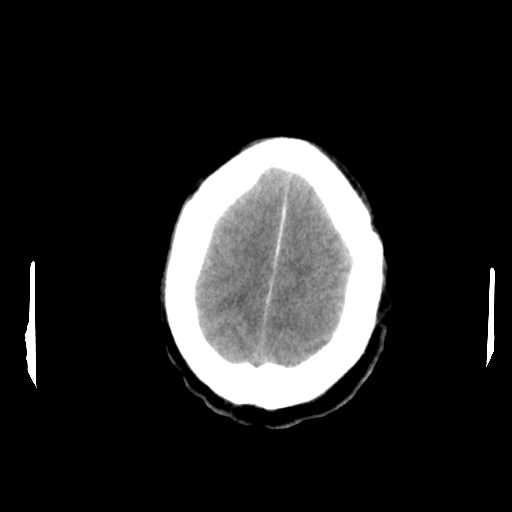}
    \end{subfigure}
    \begin{subfigure}{0.19 \textwidth}
        \includegraphics[width=\linewidth]{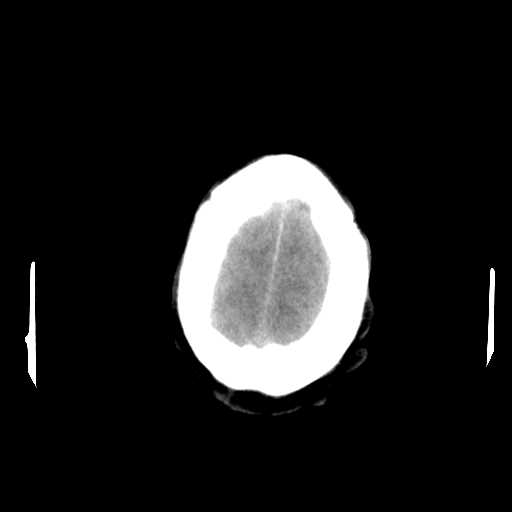}
    \end{subfigure}
    \begin{subfigure}{0.19 \textwidth}
        \includegraphics[width=\linewidth]{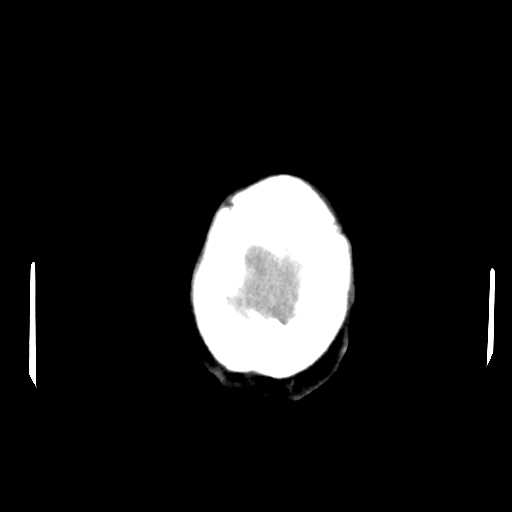}
    \end{subfigure}
    \begin{subfigure}{0.19 \textwidth}
        \includegraphics[width=\linewidth]{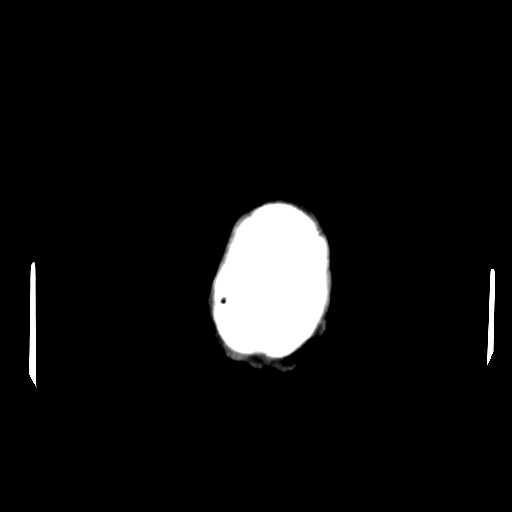}
    \end{subfigure}

    \caption{Examples of CT scan slices that were removed during preprocessing. These slices were eliminated because of lack of relevant brain tissue, or poor image quality, ensuring that only the most informative slices contribute to the classification process.}
    \label{deleted}
\end{figure*}

\section{Dataset}
\label{Dataset}

\begin{table}[t]
  \centering
  \renewcommand{\arraystretch}{2}
  \caption{Distribution of samples across different ICH subtypes.}
  \label{NumberOfSamples}
   \resizebox{\linewidth}{!}{
  \begin{tabular}{|c|c|c|c|c|c|}
    \hline
    \multirow{2}{*}{\textbf{Sample}} & \multicolumn{5}{c|}{5-Class} \\ 
    \cline{2-6} 
    & EDH & IPH & IVH & SAH & SDH \\
    \hline
    \textbf{Patient} & 89 & 421 & 154 & 66 & 118 \\ 
    \hline
    \textbf{CT scan} & 267 & 547 & 440 & 411 & 282 \\
    \hline
    \textbf{Slice} & 5945 & 9632 & 8288 & 5382 & 7394\\
    \hline
  \end{tabular}
  }
\end{table}

\subsection{Data acquisition and labeling}
The dataset used in this study was collected from two medical centers in Tehran, Iran: Rasoul Akram Hospital and Firouzabadi Hospital, over a period spanning 2018 to 2024. All CT scans were manually labeled hierarchically by two board-certified neurosurgeons to ensure accurate classification of ICH subtypes. The labeling process involved an initial annotation by one neurosurgeon, followed by validation and verification by a second expert to minimize errors and inconsistencies.
Fig. \ref{labeling} illustrates representative examples of different ICH subtypes in brain CT scans. The hemorrhagic regions are highlighted in red for better visualization. These annotated examples provide insight into the distinct imaging characteristics of each hemorrhage type, emphasizing the variability in their locations and appearances within the brain.
The study received ethical approval, and data collection was conducted in compliance with institutional review board (IRB) regulations and the Declaration of Helsinki, ensuring patient confidentiality and adherence to ethical standards.

To prevent data leakage, special attention was given to how CT scans were assigned to training and test sets. Since some patients had multiple CT scans, all scans from the same patient were entirely allocated to either the training or test dataset, ensuring subject independence in the evaluation process. Using a k-fold cross-validation approach was not workable in this scenario, as it would have resulted in imbalanced folds, where some patients’ CT scans might appear in both training and validation sets, leading to data leakage and inflated performance estimates. To ensure a robust and reliable assessment, a strict patient-level split was adopted, and the model was evaluated over five independent runs, with different random splits of the dataset in each run. In each split, 80\% of the data was allocated to the training set, while the remaining 20\% was reserved for testing. This approach provides a more reliable estimate of the model’s generalization ability while maintaining subject independence in each experiment.

Table \ref{NumberOfSamples} presents a detailed breakdown of the dataset, which comprises five ICH subtypes: EDH, IPH, IVH, SAH, and SDH. The dataset includes 848 patients, 1,947 CT scans, and 36,641 slices, with each CT scan comprising multiple slices, providing a diverse and comprehensive dataset for model training and evaluation.

\subsection{Image Preprocessing}
Slices from CT scan images that lacked relevant diagnostic information were manually removed, as illustrated in Fig. \ref{deleted}. This process excluded slices with missing brain tissue or poor image quality, ensuring that only the most informative and clinically relevant scans were used for classification. Following this initial filtering step, additional preprocessing was applied to further refine the dataset.  

To enhance the quality of input images, a background removal technique was first applied to eliminate non-homogeneous colored regions, such as white margins and unwanted edges in the CT scans. After this initial step, a binary masking approach was implemented \cite{de2017detecting} to isolate the brain region from the background. Given an input CT scan image \( I(x,y) \), a binary mask \( M(x,y) \) was generated to differentiate between the brain tissue and irrelevant regions:  

\begin{equation}
    M(x,y) =  
\begin{cases}  
1, & \text{if } I(x,y) \geq T_{\text{brain}} \\  
0, & \text{otherwise}
\end{cases} \ \ , 
\end{equation}

where \( T_{\text{brain}} \) represents an intensity threshold determined using Otsu’s thresholding method, ensuring an optimal separation between brain structures and non-brain areas. The final brain-extracted image \( I_{\text{masked}}(x,y) \) was obtained by applying the binary mask:  

\begin{equation}
    I_{\text{masked}}(x,y) = I(x,y) \cdot M(x,y)  ,
\end{equation}
With background removal and binary masking applied, the CT slices were resized from \( 528 \times 528 \) pixels to \( 224 \times 224 \) pixels using bilinear interpolation to standardize input dimensions:  
\begin{equation}
  I_{\text{resized}}(x',y') = \sum_{i=1}^{m} \sum_{j=1}^{r} I_{\text{masked}}(i,j) \cdot K(x' - i, y' - j) .  
\end{equation}
where \( m \) and \( r \) denote the height and width of the original image, and \( K(x,y) \) represents the bilinear interpolation kernel, ensuring smooth down-sampling while preserving essential anatomical features.  
By applying this hierarchical preprocessing pipeline, background removal, binary masking for brain segmentation, and resolution standardization, the model receives clean, noise-free, and size-consistent CT scan images.
\begin{figure}[ht!]
\centering
\includegraphics[scale=0.25]{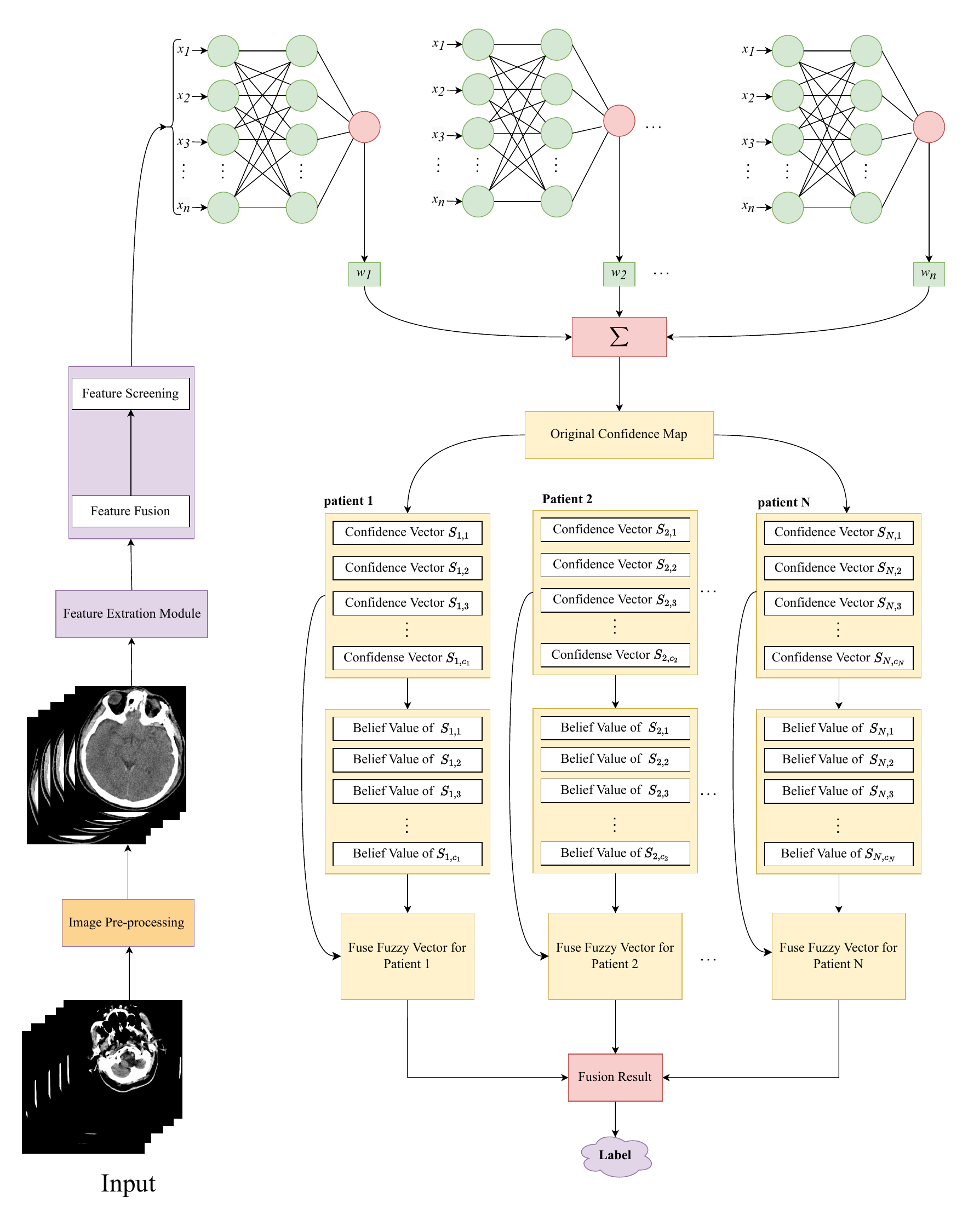}
\caption{An overview of the proposed scan-level classification framework for ICH detection.}
\label{main}
\end{figure}

\section{Methology}
\label{Methodology}
The proposed model follows a structured pipeline, beginning with image preprocessing and feature extraction techniques. The extracted features undergo feature screening and fusion, ensuring that only the most discriminative components are kept. These refined features are then processed by a boosting model, which learns high-level representations from the input CT images.  
In the next stage, multiple neural networks generate confidence vectors for each CT slice, forming the original confidence map that encapsulates slice-wise predictions. Given the hierarchical nature of CT scans, where hemorrhage characteristics may not be clearly visible in a single slice, an entropy-based belief assignment mechanism is employed. This mechanism assigns belief values to each slice based on entropy estimation, allowing the model to weigh the significance of different slices in the final decision-making process.  
A fuzzy integral-based fusion strategy aggregates information from all scan slices to achieve robust patient-level classification. This step results in fuzzy fusion vectors, which encode the overall diagnostic confidence of each patient. The final fusion result determines the ICH subtype classification, ensuring improved reliability and accuracy. The overall architecture of the proposed model is illustrated in Fig. \ref{main}, and further details of its components and fusion mechanism will be discussed in subsequent sections.

\subsection{Feature Extraction and Selection Strategy}
In the proposed model, multiple CNN-based and transformer-based architectures were independently used for feature extraction from preprocessed CT slices.
The PVT architecture, which demonstrated the best performance in feature extraction based on the achieved classification accuracy, was configured with the following hyperparameters: the patch embedding size was set to 4×4, the hierarchical feature pyramid comprised four stages with hidden dimensions of 64, 128, 320, and 512, and the number of Transformer blocks per stage was set to 2, 2, 2, and 2, respectively. A window size of 7 and head dimensions of 8 were chosen to balance local and global feature representations. The model was trained using the AdamW optimizer, with an initial learning rate of 1e-3 and a weight decay of 0.05 to enhance generalization. These hyperparameter settings ensured that the extracted features retained both high-level contextual information and fine-grained spatial details.

\begin{figure}[t]
\centering
\includegraphics[scale=0.3]{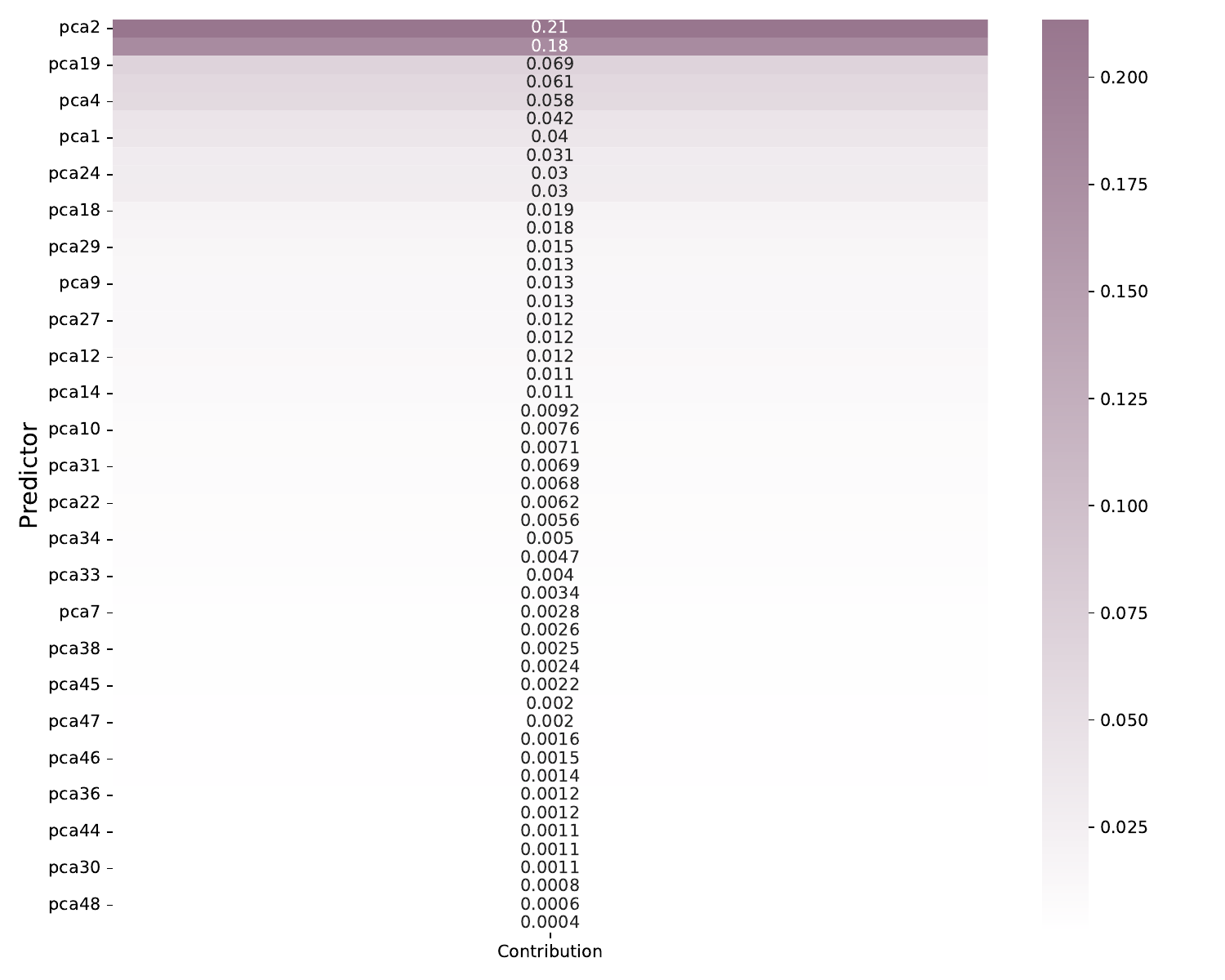}
\caption{SHAP-based feature importance ranking. The ranked components correspond to the PVT model, highlighting the most influential features contributing to the classification of ICH subtypes.}
\label{feature_imp}
\end{figure}

\begin{figure}[t]
\centering
\includegraphics[scale=0.45]{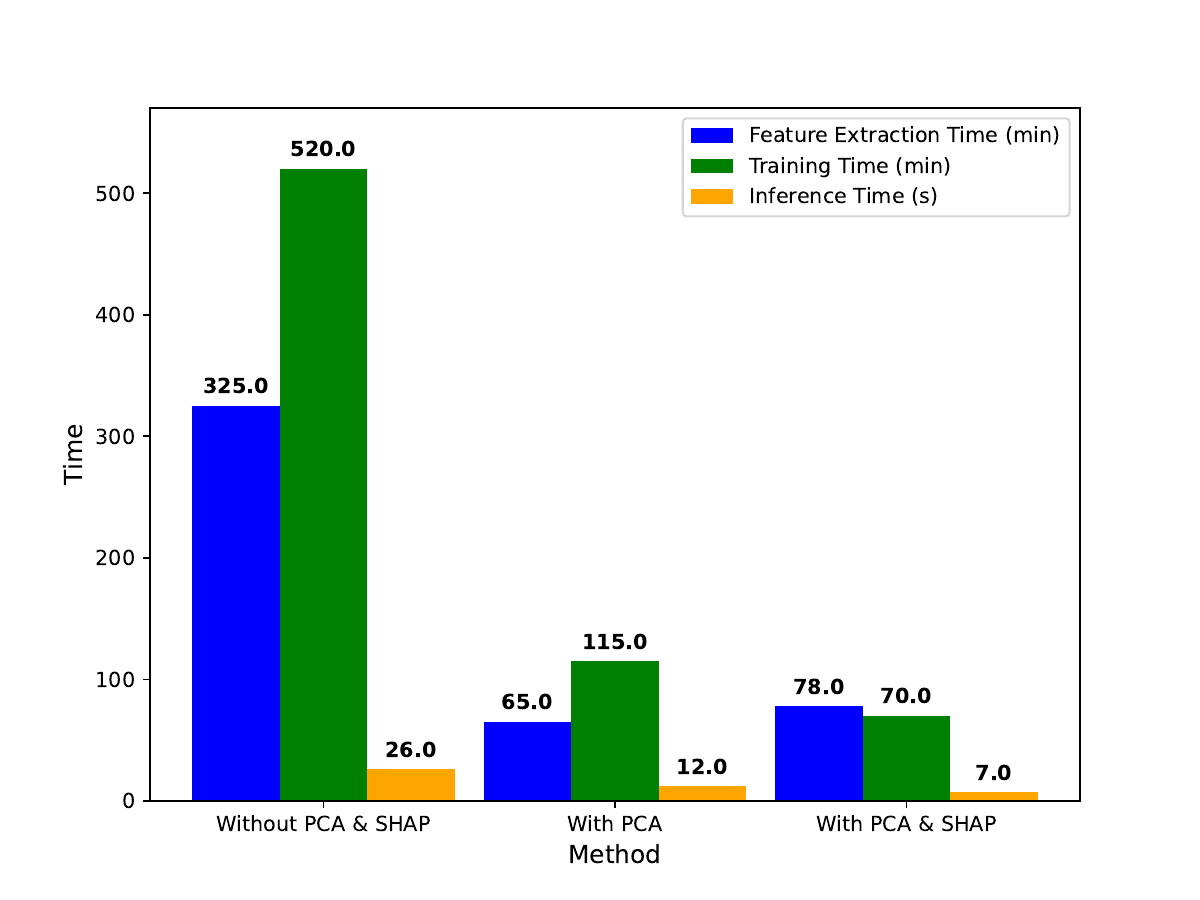}
\caption{Comparison of computational time across different processing methods in the PVT-based model.}
\label{time}
\end{figure}

To reduce dimensionality and keep the most informative components, principal component analysis (PCA) applied to the extracted features from each model. From the resulting principal components, the top 50 most significant components were kept for each feature extraction model. However, relying solely on PCA does not guarantee the most relevant components for classification, as it focuses on variance rather than feature importance for the specific ICH classification task.  
To further refine the feature space, a SHAP-based feature importance analysis was conducted. SHAP assigns an importance score to each extracted component, showing its contribution to the slice-level classification outcome. Only the components where the contribution exceeded 1\%, as illustrated in Fig. \ref{feature_imp}, were selected as the latent feature space to feed into the boosting neural network. This process ensures that the most discriminative and task-relevant features are used, improving the model’s generalization ability and reducing computational overhead. The computational efficiency of different preprocessing and feature selection techniques was evaluated by measuring the time required for feature extraction, model training, and inference. Feature extraction and training times were measured for the entire dataset, whereas inference time was computed per individual CT scan. This distinction is crucial, as inference represents the model’s real-time prediction capability, which is essential for deployment in clinical settings. The results in Fig. \ref{time} show that applying PCA significantly reduces computational costs, while integrating both PCA and SHAP further minimizes training and inference times without compromising classification performance.
\subsection{Boosting Neural Networks}
Boosting neural networks are a powerful ensemble learning technique designed to improve classification performance by sequentially training multiple weak learners and combining their outputs to form a strong predictive model. Unlike traditional neural networks that learn all patterns in a single training phase, boosting networks iteratively adjust model weights to focus on samples that are harder to classify. In each step, the model assigns greater weights to misclassified instances, thus forcing subsequent learners to prioritize these difficult cases. This process continues until a predefined number of boosting iterations is reached or the model converges. In our proposed approach, the boosting neural network comprises 10 component models, each contributing to the slice-level classification decision by capturing different aspects of the feature space. The hyperparameters of each component model are detailed in Table \ref{hyperparameter_boosting_NN}, where the optimal configurations were determined through extensive tuning to maximize classification accuracy. This architecture predicts the probability of each slice belonging to each class, generating an original confidence map that serves as the foundation for entropy estimation and subsequent decision fusion in the scan-level classification process.

\begin{table}[t]
\centering
\renewcommand{\arraystretch}{1.5}
\caption{Hyperparameter Settings for Each Component Model in the Boosting Neural Network.}
\label{hyperparameter_boosting_NN}
\resizebox{0.5\textwidth}{!}{ 
\begin{tabular}{|c|c|c|c|c|c|c|}
\hline
\multirow{2}{*}{Hidden} & \multicolumn{3}{c|}{Activation Function} & \multirow{2}{*}{Learning} & \multirow{2}{*}{Epochs} & \multirow{2}{*}{Penalty} \\ 
\cline{2-4} 
Layers & Sigmoid & Identity & Radial & Rate &  & Method \\
\hline
$L_1$ & 15 & 5 & 15 & \multirow{2}{*}{0.001} & \multirow{2}{*}{150} & \multirow{2}{*}{Squared} \\
$L_2$ & 10 & 0 & 10 &  &  & \\
\hline
\end{tabular}
}
\end{table}
\subsection{Entropy-Guided Fuzzy Integral Operator for Multi-Class CT Scan Aggregation} 
ICH classification in brain CT scans presents a significant challenge, as hemorrhagic regions may not be clearly visible in a single slice but may become apparent when analyzing multiple consecutive slices. Variations in patient positioning and scanning conditions introduce inconsistencies that can impact single-slice classification accuracy. To overcome these limitations, an entropy-aware fuzzy integral operator is employed to aggregate predictions from multiple slices into a scan-level decision while accounting for slice importance and inter-slice dependencies. This approach enhances the model’s ability to generalize across different hemorrhage types while mitigating the risk of misclassification because of slice-wise variability. 
\subsubsection{Mathematical Formulation of the Multi-Class Fuzzy Integral Operator}

Let a CT scan be represented as a set of \( n \) slices:  

\begin{equation}
\mathcal{S} = \{ s_1, s_2, ..., s_n \}.
\end{equation}

Each slice \( s_i \) is associated with a confidence vector of predicted class probabilities:

\begin{equation}
\mathbf{P}(s_i) = [P(s_i, c_1), P(s_i, c_2), ..., P(s_i, c_C)],
\end{equation}

where \( P(s_i, c_k) \) represents the probability that slice \( s_i \) belongs to class \( c_k \), and \( C \) is the number of hemorrhage subtypes (five in this case: IPH, IVH, EDH, SAH, and SDH).  
The slices are sorted based on their maximum confidence across all classes:
\begin{equation}
\begin{aligned}
\max_{k} P(s_1, c_k) &\leq \max_{k} P(s_2, c_k) \leq \dots \leq \max_{k} P(s_n, c_k), \\
\text{with} \quad P(s_0, c_k) &= 0.
\end{aligned}
\end{equation}
A fuzzy measure function \( \mu(A) \) is assigned to each subset \( A \subseteq \mathcal{S} \), where:

\begin{equation}
\mu(\emptyset) = 0, \quad \mu(\mathcal{S}) = 1, \quad \text{and} \quad \mu(A) \leq \mu(B) \quad \text{for} \quad A \subseteq B.
\end{equation}

The final scan-level classification is determined using the Choquet integral \cite{sugeno2014way}:
\begin{equation}
\mathcal{F}(\mathcal{S}, c_k) = \sum_{i=1}^{n} \left( P(s_i, c_k) - P(s_{i-1}, c_k) \right) \cdot \mu(A_i),
\end{equation}
where \( A_i = \{ s_i, s_{i+1}, ..., s_n \} \) represents the subset of remaining slices from \( s_i \) onward.
The fuzzy measure is computed iteratively:

\begin{multline}
\mu(A_i) = \mu(s_i) + \mu(s_{i+1}) + ... + \mu(s_n) \\ + \lambda \cdot \mu(s_i) \cdot \mu(s_{i+1}) \cdot ... \cdot \mu(s_n),
\end{multline}

where \( \lambda \) is the non-additivity parameter, which controls the interaction among slices and is determined by solving:

\begin{equation}
1 + \lambda = \prod_{i=1}^{n} (1 + \lambda \cdot \mu(s_i)), \quad -1 < \lambda < +\infty.
\end{equation}

If the sum of individual slice measures equals one, i.e.,

\begin{equation}
\sum_{i=1}^{n} \mu(s_i) = 1,
\end{equation}

then \( \lambda = 0 \), simplifying the aggregation process. When \( \sum_{i=1}^{n} \mu(s_i) < 1 \), the parameter lies in \( 0 < \lambda < +\infty \), while \( \sum_{i=1}^{n} \mu(s_i) > 1 \) results in \( -1 < \lambda < 0 \).

\subsubsection{Optimization of \( \lambda \) for Multi-Class CT Scan Aggregation}
For a scan with \( n \) slices, computing \( \lambda \) requires solving an \( n \)-degree nonlinear equation. Given that \( n \) varies across CT scans, each scan requires solving a unique equation, leading to significant computational complexity. Since, based on the entropy-based slice weight assignment, the condition \( \sum_{i=1}^{n} \mu(s_i) > 1 \) always holds for each CT scan, \( \lambda \) is treated as a hyperparameter and optimized using a grid search approach within the range \( -1 < \lambda < 0 \). This optimization strategy avoids solving high-order equations while preserving computational efficiency.

\subsubsection{Entropy-Based Slice Weight Assignment in Multi-Class Classification} 
Unlike traditional fuzzy integrals \cite{klement2009universal, sugeno2014way} where slice importance is determined heuristically, this approach assigns slice weights based on entropy estimation. Given the predicted probability vector for each slice \( s_i \) in scan \( j \):

\begin{equation}
\mathbf{p}_{ij} = [p_{ij,1}, p_{ij,2}, ..., p_{ij,C}],
\end{equation}

entropy is quantified as:

\begin{equation}
E(s_i) = 1 - \max_{k} P(s_i, c_k).
\end{equation}

Slices with lower entropy (i.e., higher confidence) are assigned greater importance in the aggregation process:

\begin{equation}
\mu(s_i) = 1 - E(s_i) = \max_{k} P(s_i, c_k).
\end{equation}
This weighting mechanism ensures that slices with strong diagnostic confidence contribute more significantly to the final classification decision.

\subsubsection{Final Scan-Level Decision Using the Fuzzy Integral Operator}  
After computing individual slice probabilities using a boosting neural network, the scan-level classification is determined via the weighted aggregation:

\begin{equation}
\hat{y}_j = \arg \max_{c_k} \mathcal{F}(\mathcal{S}, c_k).
\end{equation}

This method guarantees that slices with clear hemorrhage features have a higher influence, the aggregation process accounts for slice interactions to reduce noise effects, and the approach remains computationally efficient through hyperparameter tuning.
The entropy-aware fuzzy integral operator provides a robust framework for multi-class ICH classification by integrating multiple slices into a single, reliable diagnosis. By leveraging confidence-based weighting and an optimized fuzzy measure computation, this method enhances diagnostic accuracy while addressing the limitations of single-slice classification models.

\subsection{Classification and Regression-Based Metrics for Model Performance}
The performance of the five-class ICH classification model is evaluated using multiple standard classification metrics. Accuracy (\( A_c \)) measures the proportion of correctly classified instances among all instances and is computed as \( A_c = \frac{\sum_{i=1}^{C} TP_i}{N} \), where \( C \) represents the number of classes, \( TP_i \) is the number of correctly classified instances in class \( i \), and \( N \) is the total number of samples.

Macro-averaged precision (\( P_r \)) quantifies the proportion of correctly predicted positive cases out of all predicted positive cases across all classes:  
\( P_r = \frac{1}{C} \sum_{i=1}^{C} \frac{TP_i}{TP_i + FP_i} \).

Macro-averaged sensitivity (\( S_e \)) measures the proportion of actual positive cases correctly identified:  
\( S_e = \frac{1}{C} \sum_{i=1}^{C} \frac{TP_i}{TP_i + FN_i} \).

Macro-averaged specificity (\( S_p \)) evaluates the proportion of actual negative cases correctly classified:  
\( S_p = \frac{1}{C} \sum_{i=1}^{C} \frac{TN_i}{TN_i + FP_i} \).

Macro-averaged F1-score (\( F_1 \)) is the harmonic mean of precision and sensitivity:  
\( F_1 = \frac{1}{C} \sum_{i=1}^{C} 2 \times \frac{(P_r)_i \times (S_e)_i}{(P_r)_i + (S_e)_i} \).

In addition to classification performance metrics, several statistical measures are used to assess the predictive quality of the model. Generalized R-square (\( R^2_G \)) provides an estimate of how well the model explains the variance in the data and is computed as  
\( R^2_G = 1 - e^{\frac{LL_0 - LL_M}{N}} \),  
where \( LL_0 \) is the log-likelihood of the null model, \( LL_M \) is the log-likelihood of the fitted model, and \( N \) is the total number of samples.

Entropy R-square (\( R^2_E \)) evaluates the improvement of the model compared to a baseline and is defined as  
\( R^2_E = \frac{LL_M - LL_0}{LL_0} \).

Root average squared error (RASE) measures the overall deviation between predicted and actual values and is given by  
\( RASE = \sqrt{\frac{1}{N} \sum_{i=1}^{N} (\hat{y_i} - y_i)^2} \),  
where \( \hat{y_i} \) is the predicted probability for sample \( i \) and \( y_i \) is the actual observed label.

Mean absolute deviation (MAD) computes the average absolute error in predictions:  
\( MAD = \frac{1}{N} \sum_{i=1}^{N} | \hat{y_i} - y_i | \).

Log-likelihood (LL) quantifies the probability of observing the given data under the predicted model:  
\( LL = \sum_{i=1}^{N} \left[ y_i \log(\hat{y_i}) + (1 - y_i) \log(1 - \hat{y_i}) \right] \),  
where \( \hat{y_i} \) represents the predicted probability of the positive class for sample \( i \), and \( y_i \) is the actual class label. A higher log-likelihood indicates a better-fitting model.

These combined metrics ensure a comprehensive evaluation of the model’s classification accuracy, predictive reliability, and clinical applicability.
\begin{table}[t]
    \centering
    \caption{ Regression-Based Metrics for Test Data. The table presents the \( R^2_G \), \( R^2_E \), RASE, MAD, and LL for different DL models in ICH subtypes classification.}
    \label{regression}
    \resizebox{\columnwidth}{!}{ 
    \begin{tabular}{|l|c c c c c|}
    \hline
    \textbf{Model} & \multicolumn{5}{c|}{\textbf{Test Set}} \\
    \cline{2-6}
    & \textbf{$R^2_G$} & \textbf{$R^2_E$} & RASE & MAD & LL \\
    \hline
    VGG19      & 0.884 & 0.592 & 0.466 & 0.358 & -7900 \\
    ResNet50   & 0.835 & 0.473 & 0.649 & 0.501 & -12729 \\
    PVT    & 0.880 & 0.585 & 0.477 & 0.399 & -8037 \\
    CCA     & 0.849 & 0.529 & 0.516 & 0.463 & -9117 \\
    Deit        & 0.614 & 0.279 & 0.613 & 0.539 & -13960 \\
    DenseNet121     & 0.895 & 0.615 & 0.455 & 0.357 & -7454 \\
    EfficientNetB6     & 0.766 & 0.418 & 0.571 & 0.511 & -11274 \\
    Incep-ResNet     & 0.724 & 0.373 & 0.596 & 0.540 & -12144 \\
    Xception    & 0.792 & 0.448 & 0.551 & 0.478 & -10691 \\
    InceptionV3   & 0.706 & 0.355 & 0.577 & 0.485 & -12487 \\
    MobileNetV2      & 0.814 & 0.476 & 0.536 & 0.459 & -10136 \\
    Cross Attention   & 0.700 & 0.350 & 0.616 & 0.578 & -12589 \\
    ResNet152   & 0.846 & 0.525 & 0.517 & 0.453 & -9194 \\
    VGG16      & 0.883 & 0.589 & 0.477 & 0.401 & -7947 \\
    \hline
    \end{tabular}
    }
\end{table}

\begin{table}[t]
    \centering
    \caption{ Classification Metrics for Slice-Level Test Data. The table presents the \( A_C \), \( P_r \), \( S_e \), \( S_p \), and \( F_1 \) for different DL models in ICH subtypes classification.}
    \label{classification}
    \resizebox{\columnwidth}{!}{ 
    \begin{tabular}{|c|c|c|c|c|c|}
    \hline
    \textbf{Model} & \textbf{$A_c$} & \textbf{$P_r$} & \textbf{$S_e$} & \textbf{$S_p$} & \textbf{$F_1$} \\ \hline
    VGG19    & 0.709    & 0.705     & 0.703       & 0.926       & 0.704    \\ 
    ResNet50  & 0.689    & 0.685     & 0.675       & 0.921       & 0.676    \\ 
    PVT  & 0.707    & 0.704     & 0.697       & 0.926       & 0.699    \\ 
    CCA   & 0.694    & 0.690     & 0.688       & 0.922       & 0.689    \\ 
    Deit      & 0.527    & 0.512     & 0.511       & 0.881       & 0.511    \\ 
    DenseNet121    & 0.728    & 0.723     & 0.721       & 0.931       & 0.722    \\ 
    EfficientNetB6   & 0.599    & 0.587     & 0.582       & 0.899       & 0.582    \\ 
    Incep-ResNet   & 0.587    & 0.577     & 0.569       & 0.895       & 0.571    \\ 
    Xception  & 0.588    & 0.581     & 0.578       & 0.896       & 0.578    \\ 
    InceptionV3 & 0.577    & 0.562     & 0.560       & 0.893       & 0.561    \\ 
    MobileNetV2   & 0.632    & 0.617     & 0.615       & 0.907       & 0.615    \\ 
    Cross Attention & 0.557    & 0.542     & 0.534       & 0.887       & 0.532    \\ 
    ResNet152 & 0.671    & 0.662     & 0.656       & 0.917       & 0.656    \\ 
    VGG16    & 0.712    & 0.707     & 0.703       & 0.927       & 0.704    \\ \hline
    \end{tabular}
    }  
\end{table}

\section{Findings and Analysis}
\label{Findings}
In this section, we analyze the results in detail and present various performance metrics to evaluate the effectiveness of the proposed algorithm.

Tables \ref{regression} and \ref{classification} present the slice-level performance of various DL models for ICH classification. Table \ref{regression} evaluates regression-based metrics, while Table \ref{classification} reports classification metrics.  
The results indicate that VGG16, VGG19, and DenseNet121 achieve strong predictive performance, demonstrating high \( R_G^2 \) and \( R_E^2 \) values along with competitive classification metrics. The PVT model further enhances performance by improving feature interaction, while the co-scale convolutional attention (CCA) model also performs well, showing a strong balance across all evaluation criteria. In contrast, the data-efficient image transformer (DeiT) underperforms, suggesting that standard transformer architectures may be less effective than CNN-based models for slice-level ICH classification.   

\begin{figure}[t]
\centering
\includegraphics[scale=0.45]{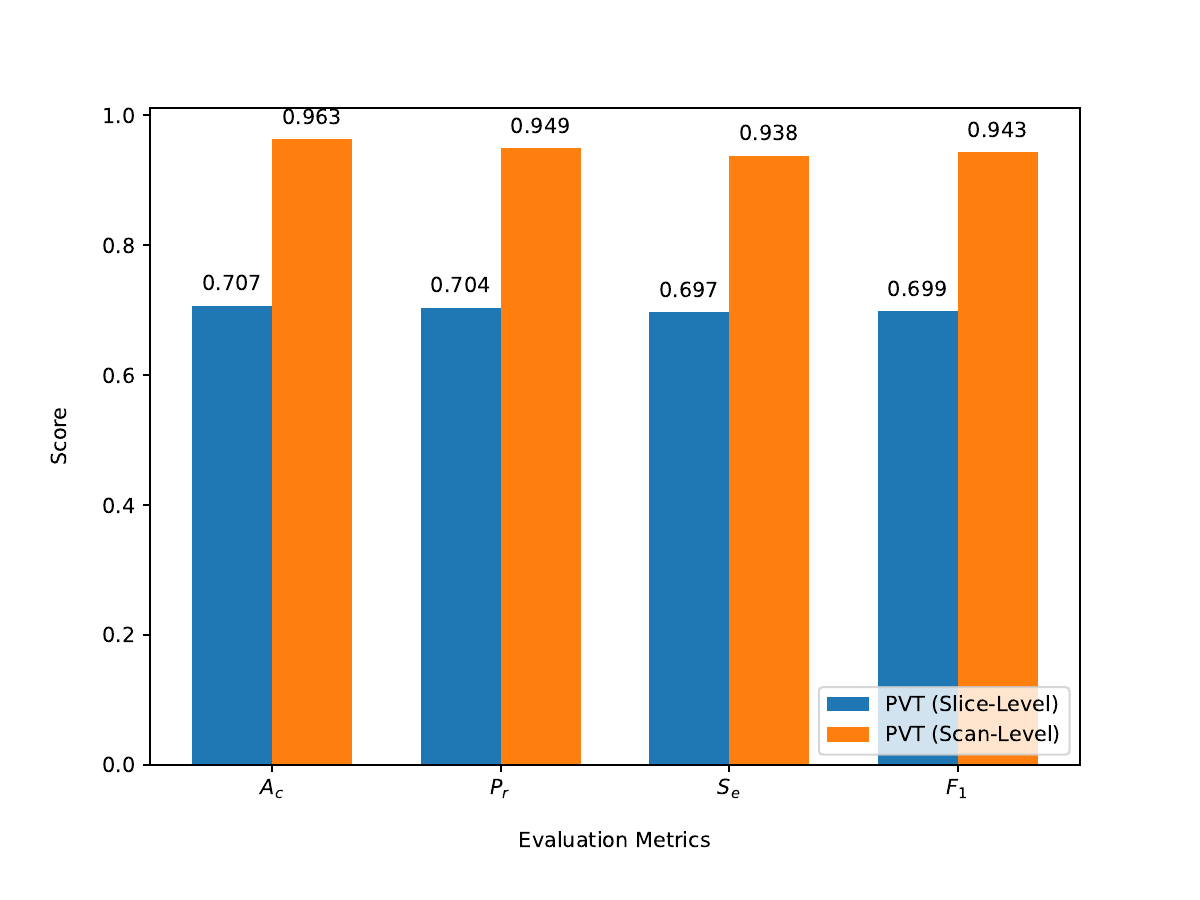}
\caption{Comparison of PVT model performance at the slice-level and scan-level classification.}
\label{slice_scan}
\end{figure}

Fig. \ref{slice_scan} compares the performance of the PVT model at the slice-level and scan-level. The slice-level classification is performed on individual CT slices, whereas the scan-level classification uses an entropy-aware fuzzy integral operator to aggregate predictions across multiple slices belonging to the same scan. The scan-level fusion significantly improves all evaluation metrics, highlighting the advantage of incorporating multi-slice information to mitigate inconsistencies that may arise from analyzing slices in isolation. This enhancement is crucial for ICH classification, where different slices within a scan may contain varying degrees of diagnostic information.

\begin{figure}[t!]
\centering
\includegraphics[scale=0.37]{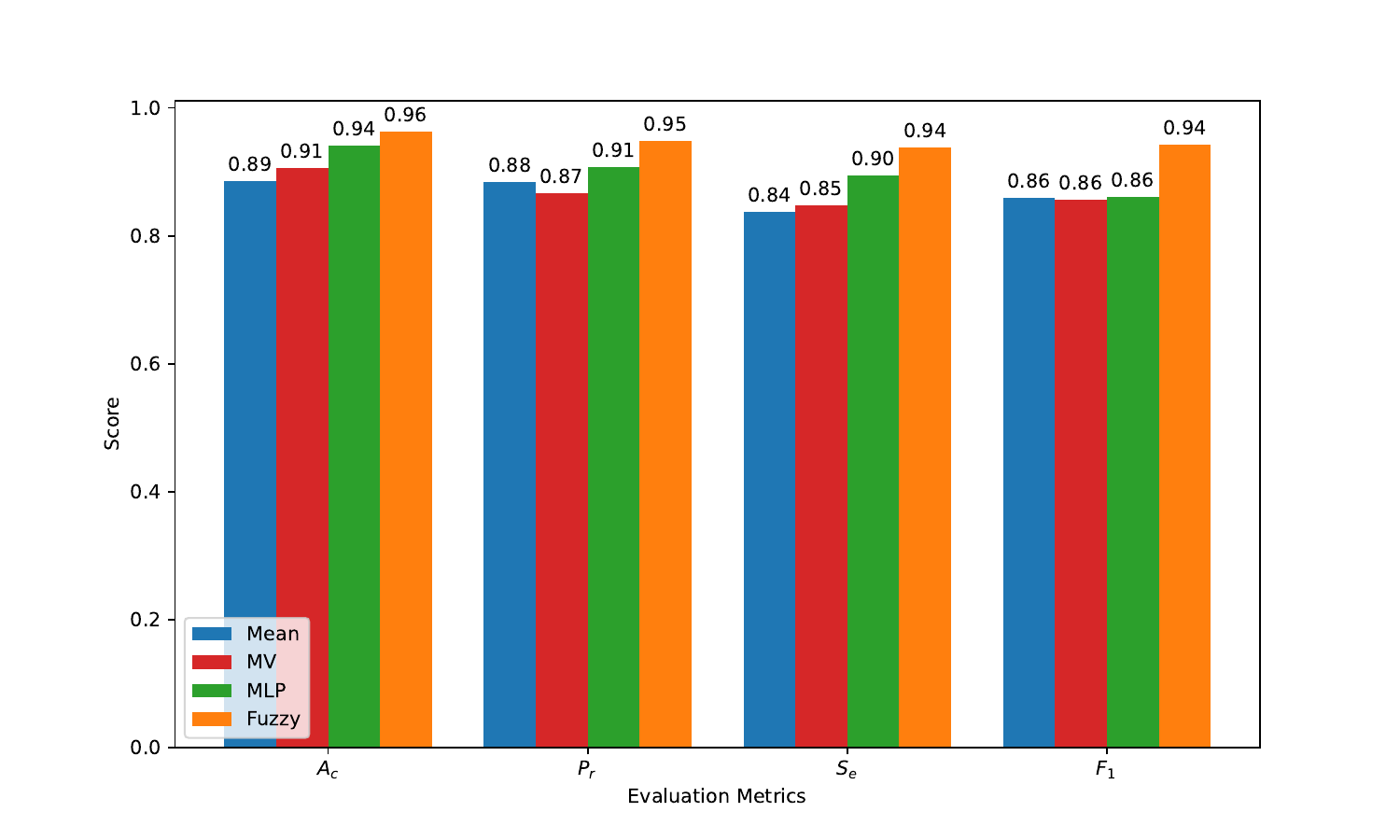}
\caption{Comparison of different scan-level decision fusion strategies using features extracted by the PVT model.}
\label{Mean_MV_MLP_Fuzzy}
\end{figure}

Fig. \ref{Mean_MV_MLP_Fuzzy} compares the performance of four different scan-level decision fusion strategies applied to PVT-extracted features. While mean aggregation and majority voting (MV) provide simple fusion mechanisms, they may cannot account for varying diagnostic confidence across slices. The multi-layer perceptron (MLP)-based approach improves performance by learning a nonlinear fusion function. In contrast, the entropy-based fuzzy integral method achieves the highest performance by dynamically weighting slice contributions based on their diagnostic confidence, ensuring a more reliable scan-level classification. These findings emphasize the importance of adaptive decision fusion strategies to enhance model robustness in ICH classification.

\begin{figure}[t]
\centering
\includegraphics[scale=0.45]{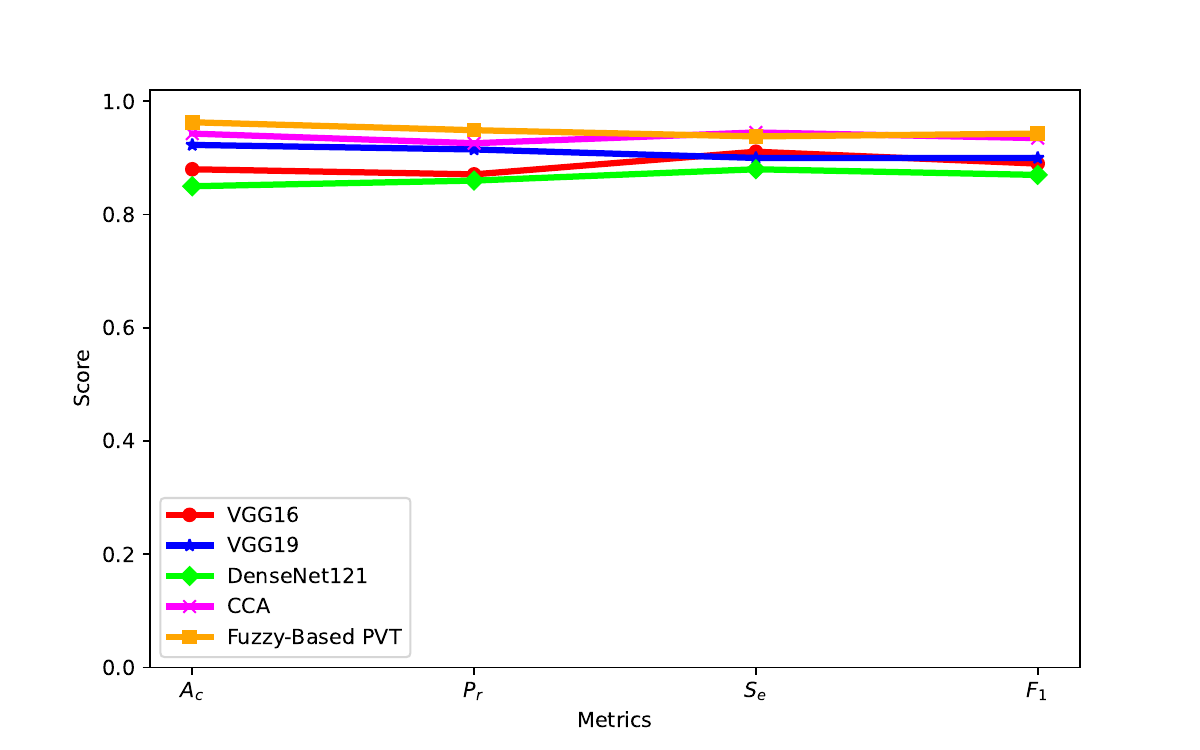}
\caption{Comparison of different DL models for scan-level classification after slice-wise fusion using the entropy-based fuzzy integral operator. }
\label{VGG_Vision}
\end{figure}

Fig. \ref{VGG_Vision} compares the scan-level classification performance of different DL architectures, including VGG16, VGG19, DenseNet121, and co-scale convolutional attention (CCA). Each model’s outputs were aggregated using the entropy-based fuzzy integral operator to ensure a comprehensive scan-level decision. These models were selected as they showed the best slice-level classification performance compared with other tested architectures. While traditional convolutional networks such as VGG and DenseNet achieve competitive performance, they exhibit limitations in capturing long-range dependencies within CT scans. The CCA model improves feature interactions through attention mechanisms, yet it is surpassed by the fuzzy-based PVT model, which leverages hierarchical spatial representations and entropy-aware fusion. These findings highlight the effectiveness of incorporating vision transformers alongside fuzzy-based decision aggregation for more reliable ICH classification.

\section{Discussion}
\label{Discussion}
In this section, we discuss the findings in relation to previous studies and analyze the implications of the proposed approach compared to the existing methodologies.

The proposed model in \cite{chen2024efficient} employed a combination of depthwise separable convolutions and a multi-receptive field mechanism. Gradient-weighted class activation mapping (Grad-CAM) was used to visualize the model’s focus areas during classification. This technique provided insights into the regions of the CT scans that contributed most to the model’s predictions, enhancing interpretability.

The study \cite{he2024deep} introduced a deep multiscale convolutional feature learning framework for the classification and weakly supervised localization of ICH based on brain CT scans. The proposed model incorporated several key modules, including a window setting optimization module, multiscale feature fusion, and weakly supervised lesion localization. The model employed VGG16 as its backbone, enhanced by attention mechanisms.

Ragab et al. \cite{ragab2023political} proposed the political optimizer with DL framework for the diagnosis of ICH. The approach begins by applying bilateral filtering to the CT images for noise reduction while maintaining edge details. Feature extraction is then performed using the Faster SqueezeNet model to identify significant features within the images. Last, classification is achieved through a denoising autoencoder model.

This study \cite{negm2023intracranial} integrated computer vision and ensemble learning for automated ICH classification. It used a multi-head attention-based CNN for feature extraction. For classification, a majority voting ensemble method was employed, comprising recurrent neural networks, bi-directional long short-term memory, and extreme learning machine-stacked autoencoder.

Malik et al. \cite{malik2023computational} presented a novel neuro-fuzzy-rough DL model for the classification of brain hemorrhages from CT images. The model combined three powerful techniques: neural networks, fuzzy logic, and rough set theory to handle the challenges of data uncertainty and overlapping features found in biomedical imaging. Tested on brain hemorrhage datasets, the model achieved a high accuracy of 96.77\% in training and 94.52\% in testing.

This study \cite{gudadhe2023classification} focused on classifying ICH in CT images using texture analysis and ensemble-based ML algorithms. Three texture-based feature extraction methods were applied: local binary pattern, local ternary pattern, and weber local descriptor. The features extracted from these methods were used to create feature vectors, and various classifiers were trained on these features.

Zhang et al. \cite{zhang2022weakly} introduced a weakly supervised guided soft attention network for the classification of ICH from brain CT scans. The method addressed the challenge of noisy and non-diagnostic features in CT images by using weak segmentation labels to guide a soft attention mechanism, focusing the model on diagnostically relevant regions.

Kyung et al. \cite{kyung2022improved} proposed a novel SMART-Net framework for the classification and segmentation of ICH from non-contrast head CT (NCCT) images. The method employed multi-task learning, where the model simultaneously performed classification, segmentation, and image reconstruction as pretext tasks to improve feature extraction and generalizability.

Lee et al. \cite{lee2019explainable} introduced an explainable algorithm for the detection of acute ICH using small datasets of brain CT scans. The proposed model combined multiple pre-trained deep convolutional neural networks, including VGG16, ResNet-50, Inception-v3, and Inception-ResNet-v2, to develop a system that achieved high diagnostic accuracy.

The paper \cite{nizarudeen2024comparative} focused on advanced methods for detecting and grading ICH using DL architectures. The research evaluated three specific architectures based on residual networks (HResNet, HResNet-SE, HRaNet). The research also considered practical aspects, such as the prediction time and the number of parameters in the models, providing insights into their feasibility for clinical integration. HRaNet achieved an AUC exceeding 0.96, showing high diagnostic performance.

Table \ref{disscusion_table} provides a comparative analysis of our proposed approach to existing DL models for ICH classification. While prior studies have showed strong performance, our model distinguishes itself through its superior classification accuracy and integrating a fuzzy fusion mechanism, which enhances decision reliability at the scan level. Using the PVT with fuzzy fusion enables improved feature extraction and robust aggregation of multi-slice predictions, leading to better overall performance. To ensure the reliability of our evaluation, the model’s performance was assessed over five independent runs, and the mean and standard deviation of key performance metrics were reported. This approach accounts for the variability introduced by different training and testing splits, providing a more robust estimate of the model’s generalization ability. Including standard deviation values highlights the model’s consistency and stability across multiple runs, ensuring that the reported improvements are not because of random fluctuations but show genuine performance gains.

\begin{table}[t]
\caption{Comparative analysis of the proposed method against existing approaches in ICH classification.}
\label{disscusion_table}
\resizebox{\linewidth}{!}{%
\begin{tabular}{|l|l|l|l|l|l|l|l|}
\hline
Work &
  Year & Dataset &
  \begin{tabular}[c]{@{}l@{}}Number of \\ Samples\end{tabular} &
  Objective &
  Deep Learning Model &
  \begin{tabular}[c]{@{}l@{}}Validation\\  Strategy\end{tabular} &
  Performance \\ \hline

{\cite{ozaltin2023classification}} &
  2023 & Public &
  815 slices &
  4-class &
  \begin{tabular}[c]{@{}l@{}}OzNet and fully connected network\end{tabular} &
  10-fold &
  \begin{tabular}[c]{@{}l@{}}AUC=0.99, $A_c$= 0.93\end{tabular} \\ \hline

{\cite{asif2023intracranial}} &
  2023 & Public &
  500 scans &
  6-class &
  \begin{tabular}[c]{@{}l@{}}ResNet101-InceptionV4-LGBM\end{tabular} &
  Single-fold &
  \begin{tabular}[c]{@{}l@{}}$A_c$= 0.95, $S_e$=0.95,\\ $S_p$= 0.94\end{tabular} \\ \hline

  {\cite{arman2023intracranial}} &
  2023 & Private &
  1 074 271 slices &
  4-class &
  \begin{tabular}[c]{@{}l@{}}DenseNet with \\ bayesian optimization \end{tabular} &
  Single-fold &
  \begin{tabular}[c]{@{}l@{}}$A_c$= 0.948,  $P_r$= 0.854\end{tabular} \\ \hline
  
  {\cite{ragab2023political}} &
  2023 & Private &
  341 scans &
  5-class &
  \begin{tabular}[c]{@{}l@{}}Faster SqueezeNet \\with denoising autoencoder \end{tabular} &
  Single-fold &
  \begin{tabular}[c]{@{}l@{}}$A_c$= 0.984, $S_e$ = 0.908, \\  $P_r$= 0.948, $F_1$= 0.921\end{tabular} \\ \hline

{\cite{negm2023intracranial}} &
  2023 & Private &
  341 scans &
  5-class &
  \begin{tabular}[c]{@{}l@{}}Majority voting (RNN, BiLstm, Stacked autoencoder) \end{tabular} &
  Single-fold &
  \begin{tabular}[c]{@{}l@{}}$A_c$= 0.98, $S_e$ = 0.93,\\ $P_r$= 0.98,  $F_1$= 0.95\end{tabular} \\ \hline

{\cite{chen2024efficient}} &
  2024 & Public &
  19 503 scans &
  6-class &
  \begin{tabular}[c]{@{}l@{}} model combining depthwise separable\\ convolutions with a multi-receptive field mechanism\end{tabular} &
  5-fold &
  \begin{tabular}[c]{@{}l@{}}AUC=0.965\end{tabular} \\ \hline
  
{\cite{he2024deep}} &
  2024 & Public &
  750 scans &
  6-class &
  \begin{tabular}[c]{@{}l@{}}VGG16 with attentional \\ fusion mechanism\end{tabular} &
  Single-fold &
  AUC = 0.973 \\ \hline
{\cite{d2024accuracy}} &
  2024 & Private &
  405 scans &
  5-class &
  \begin{tabular}[c]{@{}l@{}}DenseNet-Unet\end{tabular} &
  Single-fold &
  \begin{tabular}[c]{@{}l@{}}$A_c$= 0.91,  $S_e$= 0.9, \\ $S_p$= 0.94\end{tabular} \\ \hline

{\cite{sindhura2024fully}} &
  2024 & Public &
  8 652 scans &
  5-class &
  CNN-RNN (cascade)-UNet&
  Single-fold &
  \begin{tabular}[c]{@{}l@{}}$S_e$= 0.95, $A_c$ = 0.93, \\   $S_p$= 0.97\end{tabular} \\ \hline

\begin{tabular}[c]{@{}l@{}}Our \\ work\end{tabular}  &
  2025 & Private &
  1 947 scans &
  5-class &
  PVT model with fuzzy fusion &
  Single-fold (5 runs) &
  \begin{tabular}[c]{@{}l@{}}$S_e$= 0.938 $\pm$ 0.005,\\ $A_c$ = 0.963 $\pm$ 0.004, \\   
  $P_r$= 0.949 $\pm$ 0.003,\\ $F_1$ = 0.943 $\pm$ 0.004\end{tabular} \\ \hline
\end{tabular}
}
\end{table}

\subsection{Potential Directions for Future Research}
\label{FutureWork}
In future studies, we aim to investigate:
\begin{itemize}
    \item Multi-Modal Data Integration: Incorporating additional imaging modalities such as MRI scans or clinical data (e.g., patient history, laboratory results) could enhance classification accuracy by providing complementary information beyond CT scans.
    \item  Self-Supervised Learning for Feature Extraction: Implementing self-supervised or contrastive learning techniques could help the model learn more generalizable representations from unlabeled data, reducing dependency on manually annotated datasets.
    \item Uncertainty-Aware Decision Fusion Enhancements: Extending the fuzzy integral operator with adaptive uncertainty quantification methods could improve decision fusion across slices, particularly in ambiguous or borderline cases.
    \item Real-Time Deployment and Optimization: Developing efficient lightweight models optimized for real-time clinical deployment on edge devices or cloud-based platforms could facilitate faster and more accessible ICH classification.
    \item Explainability and Clinical Validation: Further integrating explainability techniques such as Grad-CAM or SHAP visualizations could improve model interpretability, enabling better trust and adoption by radiologists and clinicians. Conducting large-scale multi-center validation would ensure generalizability across different populations and imaging protocols.
\end{itemize}

\section{Conclusion}
\label{Conclusion}
Accurate classification of ICH is crucial because of its high mortality rate and significant impact on patient outcomes, directly influencing treatment decisions and recovery. This study introduced a PVT-based framework for ICH subtype classification, leveraging hierarchical attention mechanisms to extract spatial features from CT scans. To enhance model efficiency and interpretability, SHAP-based feature selection was applied, selecting the most discriminative components as the latent feature space for a boosting neural network. An entropy-aware fuzzy integral operator was employed to aggregate slice-level predictions into a robust scan-level decision.
Comparative evaluations showed that the proposed approach outperforms existing DL models, achieving higher accuracy and reliability in multi-class ICH classification. Additionally, by reducing computational costs, this method enables feasibility for deployment in clinical centers with limited resources, making automated ICH diagnosis more accessible. By integrating transformer-based feature extraction, feature selection, and entropy-guided decision fusion, this framework offers a scalable and efficient solution for real-world medical applications.

\bibliography{references}

@article{ozaltin2023classification,
  title={Classification of brain hemorrhage computed tomography images using OzNet hybrid algorithm},
  author={Ozaltin, Oznur and Coskun, Orhan and Yeniay, Ozgur and Subasi, Abdulhamit},
  journal={International Journal of Imaging Systems and Technology},
  volume={33},
  number={1},
  pages={69--91},
  year={2023},
  publisher={Wiley Online Library}
}

@article{asif2023intracranial,
  title={Intracranial hemorrhage detection using parallel deep convolutional models and boosting mechanism},
  author={Asif, Muhammad and Shah, Munam Ali and Khattak, Hasan Ali and Mussadiq, Shafaq and Ahmed, Ejaz and Nasr, Emad Abouel and Rauf, Hafiz Tayyab},
  journal={Diagnostics},
  volume={13},
  number={4},
  pages={652},
  year={2023},
  publisher={MDPI}
}

@article{he2024deep,
  title={Deep multiscale convolutional feature learning for intracranial hemorrhage classification and weakly supervised localization},
  author={He, Bishi and Xu, Zhe and Zhou, Dong and Zhang, Lei},
  journal={Heliyon},
  volume={10},
  number={9},
  year={2024},
  publisher={Elsevier}
}

@article{d2024accuracy,
  title={Accuracy and time efficiency of a novel deep learning algorithm for Intracranial Hemorrhage detection in CT Scans},
  author={D’Angelo, Tommaso and Bucolo, Giuseppe M and Kamareddine, Tarek and Yel, Ibrahim and Koch, Vitali and Gruenewald, Leon D and Martin, Simon and Alizadeh, Leona S and Mazziotti, Silvio and Blandino, Alfredo and others},
  journal={La radiologia medica},
  pages={1--8},
  year={2024},
  publisher={Springer}
}

@article{ragab2023political,
  title={Political Optimizer with Deep Learning Based Diagnosis for Intracranial Hemorrhage Detection},
  author={Ragab, Mahmoud and Salama, Reda and Alotaibi, Fahd S and Abdushkour, Hesham A and Alzahrani, Ibrahim R},
  journal={IEEE Access},
  year={2023},
  publisher={IEEE}
}

@article{arman2023intracranial,
  title={Intracranial hemorrhage classification from ct scan using deep learning and bayesian optimization},
  author={Arman, Shifat E and Rahman, Sayed Saminur and Irtisam, Niloy and Deowan, Shamim Ahmed and Islam, Md Ariful and Sakib, Saadman and Hasan, Mehedi},
  journal={IEEE Access},
  year={2023},
  publisher={IEEE}
}

@article{negm2023intracranial,
  title={Intracranial Haemorrhage Diagnosis Using Willow Catkin Optimization With Voting Ensemble Deep Learning on CT Brain Imaging},
  author={Negm, Noha and Aldehim, Ghadah and Nafie, Faisal Mohammed and Marzouk, Radwa and Assiri, Mohammed and Alsaid, Mohamed Ibrahim and Drar, Suhanda and Abdelbagi, Sitelbanat},
  journal={IEEE Access},
  year={2023},
  publisher={IEEE}
}

@article{chen2024efficient,
  title={An efficient deep neural network for automatic classification of acute intracranial hemorrhages in brain CT scans},
  author={Chen, Yu-Ruei and Chen, Chih-Chieh and Kuo, Chang-Fu and Lin, Ching-Heng},
  journal={Computers in Biology and Medicine},
  volume={176},
  pages={108587},
  year={2024},
  publisher={Elsevier}
}

@article{nizarudeen2024comparative,
  title={Comparative analysis of ResNet, ResNet-SE, and attention-based RaNet for hemorrhage classification in CT images using deep learning},
  author={Nizarudeen, Shanu and Shanmughavel, Ganesh Ramaswamy},
  journal={Biomedical Signal Processing and Control},
  volume={88},
  pages={105672},
  year={2024},
  publisher={Elsevier}
}

@article{sindhura2024fully,
  title={Fully automated sinogram-based deep learning model for detection and classification of intracranial hemorrhage},
  author={Sindhura, Chitimireddy and Al Fahim, Mohammad and Yalavarthy, Phaneendra K and Gorthi, Subrahmanyam},
  journal={Medical Physics},
  volume={51},
  number={3},
  pages={1944--1956},
  year={2024},
  publisher={Wiley Online Library}
}

@article{malik2023computational,
  title={A Computational Deep Fuzzy Network-Based Neuroimaging Analysis for Brain Hemorrhage Classification},
  author={Malik, Payal and Vidyarthi, Ankit},
  journal={IEEE Journal of Biomedical and Health Informatics},
  year={2023},
  publisher={IEEE}
}

@article{gudadhe2023classification,
  title={Classification of intracranial hemorrhage CT images based on texture analysis using ensemble-based machine learning algorithms: A comparative study},
  author={Gudadhe, Santwana S and Thakare, Anuradha D and Oliva, Diego},
  journal={Biomedical Signal Processing and Control},
  volume={84},
  pages={104832},
  year={2023},
  publisher={Elsevier}
}

@article{zhang2022weakly,
  title={A weakly supervised-guided soft attention network for classification of intracranial hemorrhage},
  author={Zhang, Long and Miao, Wenlong and Zhu, Chuang and Wang, Yuanyuan and Luo, Yihao and Song, Ruoning and Liu, Lian and Yang, Jie},
  journal={IEEE Transactions on Cognitive and Developmental Systems},
  volume={15},
  number={1},
  pages={42--53},
  year={2022},
  publisher={IEEE}
}

@article{kyung2022improved,
  title={Improved performance and robustness of multi-task representation learning with consistency loss between pretexts for intracranial hemorrhage identification in head CT},
  author={Kyung, Sunggu and Shin, Keewon and Jeong, Hyunsu and Kim, Ki Duk and Park, Jooyoung and Cho, Kyungjin and Lee, Jeong Hyun and Hong, GilSun and Kim, Namkug},
  journal={Medical Image Analysis},
  volume={81},
  pages={102489},
  year={2022},
  publisher={Elsevier}
}

@article{lee2019explainable,
  title={An explainable deep-learning algorithm for the detection of acute intracranial haemorrhage from small datasets},
  author={Lee, Hyunkwang and Yune, Sehyo and Mansouri, Mohammad and Kim, Myeongchan and Tajmir, Shahein H and Guerrier, Claude E and Ebert, Sarah A and Pomerantz, Stuart R and Romero, Javier M and Kamalian, Shahmir and others},
  journal={Nature biomedical engineering},
  volume={3},
  number={3},
  pages={173--182},
  year={2019},
  publisher={Nature Publishing Group UK London}
}

@article{puy2023intracerebral,
  title={Intracerebral haemorrhage},
  author={Puy, Laurent and Parry-Jones, Adrian R and Sandset, Else Charlotte and Dowlatshahi, Dar and Ziai, Wendy and Cordonnier, Charlotte},
  journal={Nature Reviews Disease Primers},
  volume={9},
  number={1},
  pages={14},
  year={2023},
  publisher={Nature Publishing Group UK London}
}

@article{renedo2024burden,
  title={Burden of ischemic and hemorrhagic stroke across the US from 1990 to 2019},
  author={Renedo, Daniela and Acosta, Julian N and Leasure, Audrey C and Sharma, Richa and Krumholz, Harlan M and De Havenon, Adam and Alahdab, Fares and Aravkin, Aleksandr Y and Aryan, Zahra and B{\"a}rnighausen, Till Winfried and others},
  journal={JAMA neurology},
  volume={81},
  number={4},
  pages={394--404},
  year={2024},
  publisher={American Medical Association}
}

@article{smith2024cavernous,
  title={Cavernous malformations of the central nervous system},
  author={Smith, Edward R},
  journal={New England Journal of Medicine},
  volume={390},
  number={11},
  pages={1022--1028},
  year={2024},
  publisher={Mass Medical Soc}
}

@article{ahmed2023systematic,
  title={A systematic review on intracranial aneurysm and hemorrhage detection using machine learning and deep learning techniques},
  author={Ahmed, S Nafees and Prakasam, P},
  journal={Progress in Biophysics and Molecular Biology},
  volume={183},
  pages={1--16},
  year={2023},
  publisher={Elsevier}
}

@article{li2024code,
  title={Code ICH: a call to action},
  author={Li, Qi and Yakhkind, Aleksandra and Alexandrov, Anne W and Alexandrov, Andrei V and Anderson, Craig S and Dowlatshahi, Dar and Frontera, Jennifer A and Hemphill, J Claude and Ganti, Latha and Kellner, Chris and others},
  journal={Stroke},
  volume={55},
  number={2},
  pages={494--505},
  year={2024},
  publisher={Lippincott Williams \& Wilkins Hagerstown, MD}
}

@article{chagahi2024cardiovascular,
  title={Cardiovascular disease detection using a novel stack-based ensemble classifier with aggregation layer, DOWA operator, and feature transformation},
  author={Chagahi, Mehdi Hosseini and Dashtaki, Saeed Mohammadi and Moshiri, Behzad and Piran, MD Jalil},
  journal={Computers in Biology and Medicine},
  volume={173},
  pages={108345},
  year={2024},
  publisher={Elsevier}
}

@article{chagahi2024enhancing,
  title={Enhancing osteoporosis detection: An explainable multi-modal learning framework with feature fusion and variable clustering},
  author={Chagahi, Mehdi Hosseini and Dashtaki, Saeed Mohammadi and Delfan, Niloufar and Mohammadi, Nadia and Samari, Alireza and Moshiri, Behzad and Piran, Md Jalil and Faust, Oliver},
  journal={arXiv preprint arXiv:2411.00916},
  year={2024}
}

@article{seners2023role,
  title={Role of brain imaging in the prediction of intracerebral hemorrhage following endovascular therapy for acute stroke},
  author={Seners, Pierre and Wouters, Anke and Ma{\"\i}er, Benjamin and Boisseau, William and Gory, Benjamin and Heit, Jeremy J and Cognard, Christophe and Mazighi, Mikael and Gaudilliere, Brice and Lemmens, Robin and others},
  journal={Stroke},
  volume={54},
  number={8},
  pages={2192--2203},
  year={2023},
  publisher={Lippincott Williams \& Wilkins Hagerstown, MD}
}

@article{mansour2023artificial,
  title={Artificial intelligence with big data analytics-based brain intracranial hemorrhage e-diagnosis using CT images},
  author={Mansour, Romany F and Escorcia-Gutierrez, Jos{\'e} and Gamarra, Margarita and D{\'\i}az, Vicente Garc{\'\i}a and Gupta, Deepak and Kumar, Sachin},
  journal={Neural Computing and Applications},
  volume={35},
  number={22},
  pages={16037--16049},
  year={2023},
  publisher={Springer}
}

@article{chagahi2024ai,
  title={AI-Powered Intracranial Hemorrhage Detection: A Co-Scale Convolutional Attention Model with Uncertainty-Based Fuzzy Integral Operator and Feature Screening},
  author={Chagahi, Mehdi Hosseini and Piran, Md Jalil and Delfan, Niloufar and Moshiri, Behzad and Parikhan, Jaber Hatam},
  journal={arXiv preprint arXiv:2412.14869},
  year={2024}
}

@article{delfan2024ai,
  title={AI-Driven Non-Invasive Detection and Staging of Steatosis in Fatty Liver Disease Using a Novel Cascade Model and Information Fusion Techniques},
  author={Delfan, Niloufar and Moghadam, Pardis Ketabi and Khoshnevisan, Mohammad and Chagahi, Mehdi Hosseini and Hatami, Behzad and Asgharzadeh, Melika and Zali, Mohammadreza and Moshiri, Behzad and Moghaddam, Amin Momeni and Khalafi, Mohammad Amin and others},
  journal={arXiv preprint arXiv:2412.04884},
  year={2024}
}

@article{neethi2024comprehensive,
  title={A comprehensive review and experimental comparison of deep learning methods for automated hemorrhage detection},
  author={Neethi, AS and Kannath, Santhosh Kumar and Kumar, Adarsh Anil and Mathew, Jimson and Rajan, Jeny},
  journal={Engineering Applications of Artificial Intelligence},
  volume={133},
  pages={108192},
  year={2024},
  publisher={Elsevier}
}

@article{zhang2024deep,
  title={Deep-Learning-Based Microwave-Induced Thermoacoustic Tomography Applying Realistic Properties of Ultrasound Transducer},
  author={Zhang, Lejia and Wang, Qizhi and Zhao, Simin and Liu, Dantong and Li, Chenzhe and Wang, Baosheng and Wang, Xiong},
  journal={IEEE Transactions on Microwave Theory and Techniques},
  year={2024},
  publisher={IEEE}
}

@article{klement2009universal,
  title={A universal integral as common frame for Choquet and Sugeno integral},
  author={Klement, Erich Peter and Mesiar, Radko and Pap, Endre},
  journal={IEEE transactions on fuzzy systems},
  volume={18},
  number={1},
  pages={178--187},
  year={2009},
  publisher={IEEE}
}

@article{sugeno2014way,
  title={A way to Choquet calculus},
  author={Sugeno, Michio},
  journal={IEEE Transactions on Fuzzy Systems},
  volume={23},
  number={5},
  pages={1439--1457},
  year={2014},
  publisher={IEEE}
}

@article{iqbal2024hybrid,
  title={Hybrid parallel fuzzy CNN paradigm: Unmasking intricacies for accurate brain MRI insights},
  author={Iqbal, Saeed and Qureshi, Adnan N and Aurangzeb, Khursheed and Alhussein, Musaed and Wang, Shuihua and Anwar, Muhammad Shahid and Khan, Faheem},
  journal={IEEE Transactions on Fuzzy Systems},
  year={2024},
  publisher={IEEE}
}

@article{ding2024fmdnn,
  title={FMDNN: A fuzzy-guided multi-granular deep neural network for histopathological image classification},
  author={Ding, Weiping and Zhou, Tianyi and Huang, Jiashuang and Jiang, Shu and Hou, Tao and Lin, Chin-Teng},
  journal={IEEE Transactions on Fuzzy Systems},
  year={2024},
  publisher={IEEE}
}

@article{de2017detecting,
  title={Detecting morphological filtering of binary images},
  author={De Natale, Francesco GB and Boato, Giulia},
  journal={IEEE Transactions on Information Forensics and Security},
  volume={12},
  number={5},
  pages={1207--1217},
  year={2017},
  publisher={IEEE}
}

@article{ren2025conv,
  title={Conv-SdMLPMixer: A hybrid medical image classification network based on multi-branch CNN and multi-scale multi-dimensional MLP},
  author={Ren, Zitong and Liu, Shiwei and Wang, Liejun and Guo, Zhiqing},
  journal={Information Fusion},
  pages={102937},
  year={2025},
  publisher={Elsevier}
}

@article{bayoudh2024survey,
  title={A survey of multimodal hybrid deep learning for computer vision: Architectures, applications, trends, and challenges},
  author={Bayoudh, Khaled},
  journal={Information Fusion},
  volume={105},
  pages={102217},
  year={2024},
  publisher={Elsevier}
}

@article{wang2025cross,
  title={Cross-attention guided loss-based deep dual-branch fusion network for liver tumor classification},
  author={Wang, Rui and Shi, Xiaoshuang and Pang, Shuting and Chen, Yidi and Zhu, Xiaofeng and Wang, Wentao and Cai, Jiabin and Song, Danjun and Li, Kang},
  journal={Information Fusion},
  volume={114},
  pages={102713},
  year={2025},
  publisher={Elsevier}
}

\end{document}